\documentclass[11pt]{article}
\usepackage{moriond,epsfig}

\def\Journal#1#2#3#4{{#1} {\bf #2}, #3 (#4)}

\def\NPB{{\em Nucl. Phys.} B}
\def\PLB{{\em Phys. Lett.}  B}
\def\PRL{\em Phys. Rev. Lett.}
\def\PRD{{\em Phys. Rev.} D}

\def\be{\begin{equation}}
\def\ee{\end{equation}}
\def\bea{\begin{eqnarray}}
\def\eea{\end{eqnarray}}

\def\siml{{\ \lower-1.2pt\vbox{\hbox{\rlap{$<$}\lower6pt\vbox{\hbox{$\sim$}}}}\ }} 
\def\bfnabla{\mbox{\boldmath $\nabla$}}

\def\als{\alpha_{\rm s}}
\def\lQ{\Lambda_{\rm QCD}}

\begin{document}
\vspace*{1.1cm}
\title{NEW RESULTS ON INCLUSIVE QUARKONIUM DECAYS}

\author{ANTONIO VAIRO}

\address{Theory Division CERN, 1211 Geneva 23, Switzerland}

\maketitle\abstracts{
I review some recent progress, leading to a substantial reduction in the number 
of non-perturbative parameters, in the calculation of inclusive quarkonium decay 
widths in the framework of non-relativistic effective field theories.}

\section{Introduction}
Considered as bound states, heavy quarkonia 
(e.g. $\Upsilon$, $\chi_b$, $\psi$, $\chi_c$, ...) have two main characteristics:
1) a mass scale, which is in the perturbative regime: 
$m_b \simeq 5$ GeV, ~$m_c \simeq 1.5$ GeV; 
2) level splittings of a relative size that is typical of non-relativistic systems. 
Being non-relativistic bound states, heavy quarkonia are characterized 
by at least three hierarchically ordered energy scales:
the hard scale $m$, the momentum scale $mv$ and the bound-state energy scale 
$m v^2$, where $v \ll 1$ is the heavy-quark velocity inside the bound state. \cite{NRQCD}

The different scales entering the quarkonium dynamics may be 
systematically integrated out, leading from QCD to simpler but equivalent
effective field theories (EFTs). NRQCD is the EFT obtained by integrating out 
the hard scale $m$. \cite{NRQCD,BBL} This $m$ being larger than the scale 
of non-perturbative physics, $\lQ$, the matching to NRQCD can be done 
order by order in $\als$. Hence, the NRQCD Lagrangian can be written 
as a sum of terms like $ f_n \, O^{(d_n)}_n/m^{d-4}$, ordered in powers of $\als$ 
and $v$. More specifically, the Wilson coefficients 
$f_n$ are series in $\als(m)$ and encode the ultraviolet physics that  
has been integrated out from QCD. The operators $O^{(d_n)}_n$ of dimension $d_n$
describe the low-energy dynamics and are counted in powers of $v$.
Heavy quarkonium decays are processes that take place at an energy-transfer scale 
of the order of the heavy-quark mass. Their signature is encoded in the
imaginary part of the Wilson coefficients of the 4-fermion operators 
($O^{(d_n)}_n = \psi^\dagger K_n  \chi \chi^\dagger K^\prime_n \psi$) 
in the NRQCD Lagrangian. The NRQCD factorization formula for 
quarkonium inclusive decay widths into light hadrons (LH) reads \cite{BBL}
\begin{eqnarray}
& &\Gamma({\rm H}\to{\rm LH}) = 
\sum_n {2 \, {\rm Im} \, f_n \over m^{d_n - 4}}
\; \langle {\rm H} | \psi^\dagger K_n  \chi \chi^\dagger K^\prime_n \psi |{\rm H} \rangle.
\label{fac1}
\end{eqnarray}
The 4-fermion operators are classified with respect to their rotational 
and spin symmetry (e.g. $O(^{2S+1}S_J)$, $O(^{2S+1}P_J)$, ...) and of their 
colour content (octet, $O_8$, and singlet, $O_1$, operators). 
Singlet operator expectation values may be easily related to the square 
of the quarkonium wave functions (or derivatives of it) at the origin. 
These are unknown non-perturbative parameters and may be fitted from
experimental data, calculated on the lattice or taken from phenomenological 
potential models. Octet operator expectation values are also unknown 
non-perturbative parameters to be fitted from the experimental data, 
or to be calculated on the lattice. 

NRQCD still contains dynamical degrees of freedom associated with
energy scales larger than the ultrasoft scale $m v^2$. \cite{pNRQCD0} \footnote{ 
As a consequence the power counting of the NRQCD operators is not unique,
since they depend, in general, on the scales $mv$, $mv^2$ and $\lQ$.}
Hence, pushing further the EFT programme for non-relativistic bound states, 
further simplifications occur if we integrate out those degrees of freedom.
We call pNRQCD the resulting EFT. We will consider pNRQCD under 
the condition $\lQ \gg mv^2$. Then, two situations are possible.
First, the situation when $m v \gg \lQ \gg mv^2$.
In this case the soft scale $m v$ can be integrated out perturbatively. This leads to
an intermediate EFT that contains singlet and octet quarkonium fields 
and ultrasoft gluons as dynamical degrees of freedom.  
The octet quarkonium field and the ultrasoft gluons are eventually 
integrated out by the (non-perturbative) matching to pNRQCD. \cite{pNRQCD} 
Second, the situation when $\lQ \sim mv$. In this case the (non-perturbative) 
matching to pNRQCD has to be done in one single step. \cite{M1}
Under the circumstances that other degrees of freedom develop 
a mass gap of order $\lQ$ or that they play a negligible role, 
the quarkonium singlet field $\rm S$ remains as the only dynamical 
degree of freedom in the pNRQCD Lagrangian, which reads \cite{pNRQCD,M1,prl}
${\mathcal{L}}_{\rm pNRQCD}= 
{\rm Tr} \,\Big\{ {\rm S}^\dagger \left( i\partial_0 - {\cal H}  \right) {\rm S} \Big
\}$, ${\cal H}$ being the pNRQCD Hamiltonian, to be determined by matching pNRQCD to NRQCD.
The inclusive quarkonium decay width into light hadrons is given by 
\be
\Gamma ({\rm H}\to{\rm LH}) = - 2\, {\rm Im} \, \langle n,L,S,J| {\cal H}  |n,L,S,J \rangle,
\label{imag}
\ee
where $|n,L,S,J \rangle$ is an eigenstate of ${\cal H}$ with the quantum numbers 
of the quarkonium state $\rm H$. From the matching we obtain schematically:
\be
{\rm Im} \, {\cal H} = \delta^3({\bf r}) 
\sum_n {{\rm Im} \, f_n \over m^{d_n - 4}} {\cal A}_n
+ \{ \delta^3({\bf r}), \Delta \}  
\sum_n {{\rm Im} \, f_n \over m^{d_n - 4}} {\cal B}_n
+ {\bfnabla}^i\delta^3({\bf r}){\bfnabla}^j
\sum_n {{\rm Im} \, f_n \over m^{d_n - 4}} {\cal C}_n^{ij}
+ \dots,
\label{imh}
\ee
where $f_n$ are the matching coefficients inherited from NRQCD, 
and ${\cal A}_n$, ${\cal B}_n$, ... are non-perturbative operators, which are 
universal in the sense that they do not depend either on the heavy-quark flavour
or on the specific quantum numbers of the considered heavy-quarkonium state.
Inserting Eq. (\ref{imh}) into (\ref{imag}) and comparing with
Eq. (\ref{fac1}), we see that all NRQCD matrix elements, including the octet
ones, can be expressed through pNRQCD as products of the universal 
non-perturbative factors by the squares of the quarkonium wave functions
(or derivatives of it) at the origin. This drastically reduces the number of unknown 
non-perturbative factors to be introduced in order to describe the whole set 
of charmonium and bottomonium inclusive decays and makes new 
theoretical predictions possible.
As an example, in the following we will discuss the case of $P$-wave inclusive 
quarkonium decays into light hadrons. \cite{prl}

\section{P-wave decays in NRQCD and pNRQCD}
In NRQCD the $P$-wave inclusive decay width for the $S=0$ ($h$) and 
$S=1$ ($\chi$) quarkonium states is given at leading (non-vanishing) 
order in $v$ by: \cite{BBL}
\begin{eqnarray}
& & 
\Gamma(h  \to {\rm LH})
= {9 \; {\rm Im \,}  f_1(^1P_1)  \over \pi m^4} \;  
\Big| {R_P'} \Big|^2 
\;+\; {2 \; {\rm Im \,}  f_8(^1S_0)  \over m^2} \; 
\langle h | O_8 (^1S_0) | h \rangle, 
\label{Pnrqcd1}
\\  
& &
\Gamma(\chi_{J}  \to {\rm LH})
\; = \;
{9 \; {\rm Im \,}  f_1(^3P_J)  \over \pi m^4} \;  
\Big| {R_P'} \Big|^2 
\;+\; {2 \; {\rm Im \,}  f_8(^3S_1)  \over m^2} \; 
\langle h | O_8(^1S_0 ) | h \rangle, \quad {\rm for} \; J=0,1,2~~~~
\label{Pnrqcd2}
\end{eqnarray}
where $R_P'$ is the derivative of the $P$-wave function at the origin.
We stress that, according to the power counting of NRQCD, the 
octet  contribution $\langle h | O_8 (^1S_0) | h \rangle$ 
is as relevant as the singlet contribution; the above formula may by no means 
be used without considering it (for instance in 
extracting $\als(m)$, as can still be found in some recent literature). 
We also recall that the octet contribution reabsorbs the dependence 
on the infrared cut-off $\mu$ of the Wilson coefficients ${\rm Im \,} f_1(P)$.
Finally, we see that in NRQCD the 8 $P$-wave bottomonium states ($1P$, $2P$) 
and the 4 $P$-wave charmonium states ($1P$), which lie under threshold, depend  
at leading order in the velocity expansion on 6 non-perturbative parameters 
(3 wave functions $+$ 3 octet matrix elements).

\begin{figure}
\makebox[0cm]{\phantom b}
\put(120,0){\epsfxsize=6.5truecm \epsfbox{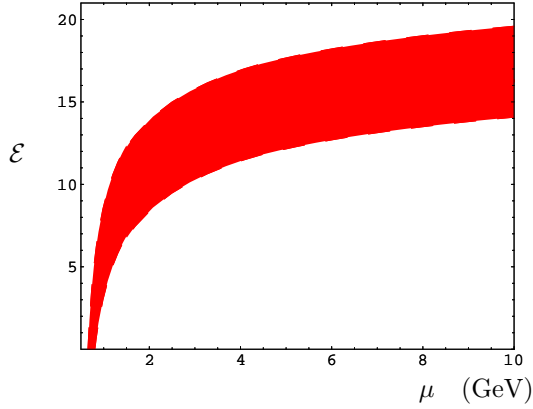}}
\put(265,-2){\small $\mu$ ~ (GeV)}
\put(110,87){\small ${\mathcal E}$}
\caption{Plot of the 1-loop RG-improved expression of 
$\mathcal E$ vs. $\mu$: $ {\mathcal{E}} (\mu)= {\mathcal{E}}(m) +
\displaystyle {96\over \beta_0}\ln {\als(m)\over \als(\mu)}$. 
${\mathcal{E}}(m)$ has been extracted from the charmonium $P$-wave data. 
The error band accounts only for the uncertainties inherited from the charmonium data.}
\label{fig1}
\end{figure}

In pNRQCD the $P$-wave inclusive decay widths are given at leading 
order in $v$ by: \cite{prl}
\begin{eqnarray}
& &
\Gamma( h  \to {\rm LH})
= {\Big| { {R_P'}} \Big|^2  \over \pi m^4} \; 
\left[ 9 \; {\rm Im \,}  f_1(^1P_1) 
\;+\; {{\rm Im \,}  f_8(^1S_0)  \over 9} \;
{\mathcal E} \right], 
\label{Ppnrqcd1}
\\  
& & 
\Gamma( \chi_{J}  \to {\rm LH})
\; = \;
{\Big| { {R_P'}} \Big|^2  \over \pi m^4} \; 
\left[ 9 \; {\rm Im \,}  f_1(^3P_J) 
\;+\; {{\rm Im \,}  f_8(^3S_1)  \over 9} \;
{\mathcal E} \right], \quad {\rm for}\; J=0,1,2
\label{Ppnrqcd2}
\end{eqnarray}
where $ {\mathcal{E}} = \displaystyle {1\over 2} \int_{0}^{\infty} d t \; t^3  
\left\langle g {\bf E}^a(t,{\bf 0}) 
\Phi_{ab}(t,0;{\bf 0}) g{\bf E}^b(0,{\bf 0}) \right\rangle$ is the universal 
non-perturbative parameter that describes $P$-wave quarkonium decays in pNRQCD. 
A plot of ${\mathcal E}$ as a function of the factorization 
scale $\mu$ is shown in Fig. \ref{fig1}.
In order to better understand the meaning of $\mathcal E$, we 
can express it in terms of the eigenstates $|n\rangle$ and the energies $E_n^{(0)}$
of the static QCD Hamiltonian: \cite{M1}
\be
{\mathcal E} =  18 \sum_{n\neq 0}
{\langle 0|g{\bf E}|n\rangle \cdot \langle n|g{\bf E}|0\rangle \over 
(E_n^{(0)} - E_0^{(0)})^4}.
\label{qmE}
\ee
Since $E_0^{(0)} \sim mv^2$ (it is the quarkonium static energy) 
and  $E_{n\neq 0}^{(0)} \sim \lQ$ (for higher gluonic
excitations we have assumed a mass gap of order $\lQ$), 
we see from Eq. (\ref{qmE}) that the correlator ${\mathcal E}$ 
resums all the corrections $(mv^2/\lQ)^n$, which one would expect 
to show up in pNRQCD after having integrated out the scale $\lQ$.

By comparing Eqs. (\ref{Pnrqcd1}) and (\ref{Pnrqcd2}) with 
Eqs. (\ref{Ppnrqcd1}) and (\ref{Ppnrqcd2}) we get at leading order in $v$ 
the relation between the octet matrix element of NRQCD and ${\mathcal E}$:
$\langle h| O_8( {}^1S_0 )| h \rangle  = \displaystyle 
\Big| { {R_P'}} \Big|^2 \; {\mathcal E}/(18 \pi m^2)$.
The quarkonium-state dependence factorizes in the pNRQCD formulas.
This allows some new predictions with respect to NRQCD, which  
are synthetized by the formula (valid at leading order in $v$): 
\be
{ \Gamma({\rm H}(^{2S+1}n{\rm{P}_J})  \rightarrow {\rm{LH}}) 
\over  \Gamma({\rm H}(^{2S'+1}n{\rm{P}_{J'}})  \rightarrow {\rm{LH}})} = 
{81\, {\rm Im}\,f_1(^{2S+1}{\rm{P}}_J)  + 
{\rm Im} \,   f_8(^{2S+1}{\rm{S}}_S) \, {\mathcal{E}}  
\over 
81 \, {\rm Im}\,f_1(^{2S'+1}{\rm{P}}_{J'})  + 
{\rm Im} \,   f_8(^{2S'+1}{\rm{S}}_{S'})\,{\mathcal{E}} },
\label{pred}
\ee
where the left-hand side is a ratio between inclusive decay widths 
of $P$-wave quarkonia with the same principal quantum number $n$ 
and the right-hand side no longer depends on $n$ and has the whole flavour 
dependence encoded in the Wilson coefficients, which are known quantities.

\begin{figure}
\makebox[0cm]{\phantom b}
\put(25,0){\epsfxsize=6.5truecm \epsfbox{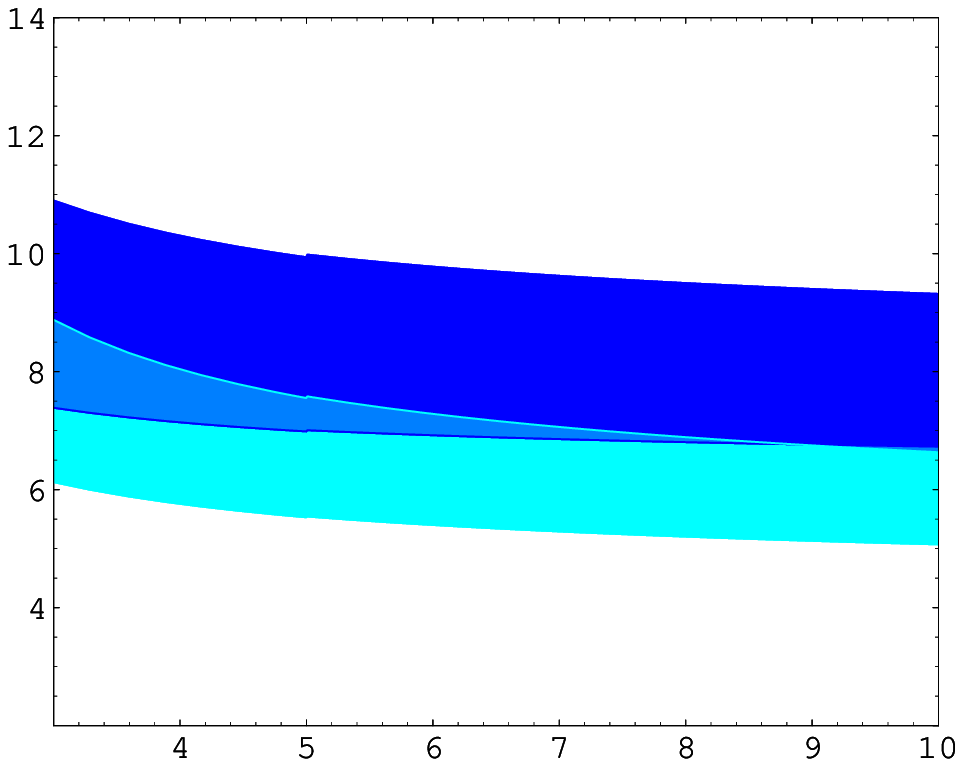}}
\put(170,-2){\small $\mu$  ~ (GeV)}
\put(0,87){\small ${\Gamma_{\chi_{b0}\rightarrow {\rm{LH}}}
               \over\Gamma_{\chi_{b1}\rightarrow {\rm{LH}}}}$}
\put(150,80){\small NLO}
\put(150,55){\small LO}
\put(265,0){\epsfxsize=6.5truecm \epsfbox{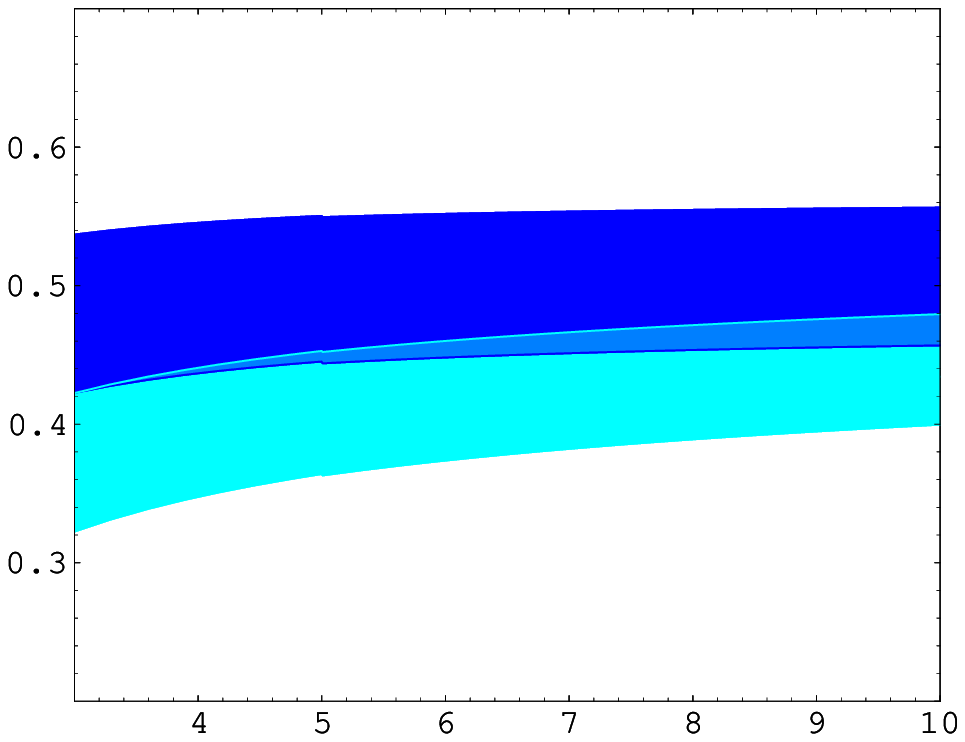}}
\put(410,-2){\small $\mu$  ~ (GeV)}
\put(237,87){\small ${\Gamma_{\chi_{b1}\rightarrow {\rm{LH}}}
                 \over\Gamma_{\chi_{b2}\rightarrow {\rm{LH}}}}$}
\put(390,95){\small NLO}
\put(390,70){\small LO }
\caption{The left-hand side of Eqs. (\ref{U01}) and (\ref{U12}) plotted 
vs. $\mu$. We have used Eq. (\ref{pred}) and the values of $\mathcal E$ from 
Fig. \ref{fig1}. The LO and NLO bands refer respectively to the Wilson  
coefficients at leading and next-to-leading order. 
\label{fig2}
}
\end{figure}

In practice, the 12 $P$-wave quarkonium states, which lie under threshold, 
depend only, in pNRQCD at leading (non-vanishing) order in the velocity
expansion, on 4 non-perturbative parameters (3 wave functions $+$ 
1 chromoelectric correlator $\mathcal E$).
The reduction by 2 in the number of unknown non-perturbative parameters 
with respect to NRQCD, allows us to formulate two new statements. 
Using the values of $\mathcal E$ from Fig. \ref{fig1} we get (at NLO):   
\be
{ \Gamma(\chi_{b0}(1P)  \rightarrow {\rm{LH}}) 
\over \Gamma(\chi_{b1}(1P)  \rightarrow {\rm{LH}})} 
=
{ \Gamma(\chi_{b0}(2P)  \rightarrow {\rm{LH}}) 
\over \Gamma(\chi_{b1}(2P)  \rightarrow {\rm{LH}})} 
= 8.0 \pm 1.3,  
\label{U01}
\ee
or alternatively 
\be
{ \Gamma(\chi_{b1}(1P)  \rightarrow {\rm{LH}}) 
\over  \Gamma(\chi_{b2}(1P)  \rightarrow {\rm{LH}})} 
=
{ \Gamma(\chi_{b1}(2P)  \rightarrow {\rm{LH}}) 
\over  \Gamma(\chi_{b2}(2P)  \rightarrow {\rm{LH}})} 
= 0.50^{+0.06}_{-0.04}.
\label{U12}
\ee
The errors refer only to the uncertainties displayed in Fig. \ref{fig1}. 
In Fig. \ref{fig2} we plot the above ratios of decay widths 
as functions of the factorization scale $\mu$. The figures show a stable result. 

\section{Conclusions}
By integrating out degrees of freedom associated with energy scales larger than
$mv^2$,  octet matrix elements in NRQCD can be written as products of 
wave functions at the origin by universal non-perturbative factors. 
This  reduces  the number of unknown matrix elements of NRQCD and 
enables  definite  predictions  for inclusive decay ratios of states with 
different flavour or principal quantum number.
As an explicit example we have shown the case of $P$-wave 
quarkonium inclusive decay into light hadrons. 
The same program may be carried out for $S$-wave decays. \cite{prep} 
At ${\cal O}(mv^5)$ the hadronic inclusive decay widths 
of the 10 $S$-wave quarkonium states, which lie under threshold, are described in NRQCD 
by about 30 non-perturbative matrix elements: 10 wave functions and 20
further matrix elements. These 20 matrix elements reduce dramatically in
pNRQCD to few universal correlators, which may be fixed 
phenomenologically or by lattice calculations.
Noting that in the hadronic/electromagnetic ratios the wave-function
dependence drops out, we expect that in the near future pNRQCD will enable QCD 
to access a large number of new results concerning heavy quarkonium inclusive and 
electromagnetic decays.

\end{document}